\documentclass[final,5p,times,twocolumn,authoryear]{elsarticle}

\usepackage[authoryear]{natbib}
\usepackage{graphicx}
\usepackage{subcaption}
\usepackage{amssymb,amsmath}
\usepackage{pdflscape}
\usepackage{adjustbox}

\usepackage{booktabs}
\usepackage{lipsum}
\usepackage{epsfig}
\usepackage{txfonts}
\usepackage{xcolor}
\usepackage{tabularx}
\usepackage{footnote}
\usepackage{soul}
\usepackage{multirow}
\usepackage{multicol}
\usepackage{mathpazo}
\usepackage{lineno,hyperref}
\usepackage{textcomp}

\usepackage[utf8]{inputenc}
\usepackage{booktabs}

\usepackage{amsthm}

\journal{High Energy Astrophysics}

\begin{document}

\begin{frontmatter}



\title{A multi-technique search for year-scale $\gamma$-ray quasi-periodic modulation in the high-redshift FSRQ PKS~2052$-$47}


%

\author{Sikandar Akbar\corref{cor1}\fnref{label1}}
\ead{darprince46@gmail.com}
\affiliation[label1]{Department of Physics, University of Kashmir, Srinagar 190006, India}
\cortext[cor1]{Corresponding author: Sikandar Akbar}






\begin{abstract}
We investigate long-timescale quasi-periodic oscillations in the $\gamma$-ray emission of the
high-redshift flat-spectrum radio quasar PKS~2052$-$47 using monthly binned \emph{Fermi}-LAT
data spanning MJD~54727.99--58507.99. To assess the statistical significance of periodic features
embedded in red-noise-dominated variability, we employ several complementary timing techniques,
including the Lomb--Scargle periodogram, weighted wavelet $Z$-transform, date-compensated discrete
Fourier transform, REDFIT assuming an AR(1) process, and damped random walk modelling.
Taken together, the analyses reveal a dominant quasi-periodic modulation on a timescale of
$\sim600$--630~d, together with a secondary longer-timescale feature near $\sim1050$--1110~d.
Monte Carlo simulations demonstrate that the shorter-period signal exceeds the highest local
confidence levels, while the longer modulation reaches $\gtrsim99$~per~cent local significance in
several tests; independent DRW-based simulations further show that both peaks rise above the
$4\sigma$ confidence envelope in the LSP analysis. Spectral-window diagnostics indicate that the
detected periodicities are not artefacts of uneven sampling. A sliding-window analysis further
shows that the QPO power varies with time, implying episodic rather than persistent modulation
across the $\sim11$~yr baseline.
We discuss possible physical interpretations in terms of geometric modulation of Doppler boosting
associated with jet precession or helical structures, accretion-driven instabilities at the jet
base, and SMBBH--induced dynamics. The presence of two timescales and their intermittent behaviour
may reflect coupled geometric processes or near-resonant modulation patterns. Our results identify
PKS~2052$-$47 as a promising case of long-timescale $\gamma$-ray QPOs and motivate future broadband
spectral modelling and coordinated multiwavelength observations to test SMBBH scenarios and to
determine the physical origin of the oscillations.
\end{abstract}



\end{frontmatter}




\section{Introduction}
\label{introduction}

The source PKS~2052$-$47 (also known as 4FGL~J2056.2$-$4714) is a $\gamma$-ray--bright flat-spectrum radio quasar located at a redshift of $z=1.489$ \citep{1984ApJ...286..498J}. Early ATCA radio imaging and \emph{Chandra} observations revealed a two-sided kiloparsec-scale jet with no extended X-ray emission \citep{Marshall_2005}. Very-long-baseline interferometry observations from the TANAMI programme show that the parsec-scale structure is extremely compact, with a high brightness-temperature core of $\sim2\times10^{12}$~K and only a very faint jet component detected at 8.4~GHz \citep{2007AAS...211.0413K,2009arXiv0912.4192B}. These properties establish PKS~2052$-$47 as a classical, strongly Doppler-boosted FSRQ.

PKS~2052$-$47 entered a phase of enhanced activity in mid--2009, when a prominent optical flare was reported by the Automatic Telescope for Optical Monitoring (ATOM; ATel~\#2158). Following a gradual brightening since July~2009, the source reached a peak magnitude of $R \simeq 15.6$ in August, representing the highest optical flux measured for this object by ATOM up to that time  \citep{2009ATel.2158....1H}.
 Shortly thereafter, the \emph{Fermi} Large Area Telescope (LAT) reported a strong enhancement in $\gamma$-ray emission above 100~MeV, establishing a close temporal association between the optical outburst and high-energy activity (ATel~\#2160; \cite{2009ATel.2160....1C}). On 2009~August~9, the source reached $F(E>100~\mathrm{MeV})=(8.7\pm1.6)\times10^{-7}\ \mathrm{ph\ cm^{-2}\ s^{-1}}$ (statistical uncertainty only), approximately a factor of four above the average level observed during the preceding week (2009~July~29--August~4), confirming PKS~2052$-$47 as a highly variable $\gamma$-ray emitter.

Motivated by this activity, a coordinated multiwavelength campaign was conducted in September 2009, covering radio, sub-millimeter, optical/UV, X-ray, and $\gamma$-ray bands. \textit{Swift}-XRT observations revealed a hard X-ray spectrum well described by an absorbed power-law model with a photon index of $\Gamma_{\rm X} \approx 1.5$, while contemporaneous \textit{Fermi}-LAT data showed a gradual rise in $\gamma$-ray flux leading into the flaring phase. The resulting broadband spectral energy distribution (SED), constructed from quasi-simultaneous observations, displayed the characteristic double-humped blazar structure. The inverse-Compton component was found to dominate over the synchrotron peak, consistent with earlier historical SEDs and indicating that high-energy emission is primarily produced through inverse-Compton scattering processes \citep{2010arXiv1001.1563C}.

Subsequent observations demonstrated that PKS~2052$-$47 has remained intermittently active throughout the \emph{Fermi} era. In March~2020, the source again exhibited enhanced $\gamma$-ray emission, with the daily averaged flux above 100~MeV reaching $F(E>100~\mathrm{MeV})\simeq1.2\times10^{-6}\ \mathrm{ph\ cm^{-2}\ s^{-1}}$, nearly an order of magnitude higher than the long-term average reported in the fourth \emph{Fermi}-LAT catalog (ATel~\#13541; \citealt{2020ATel13541....1B}). During this episode, the photon index was $\Gamma_{\gamma}=2.3\pm0.2$, consistent with the catalog value, indicating that the variability was dominated by flux changes rather than strong spectral-shape variations.

In addition to episodic flaring, PKS~2052$-$47 has been identified as a candidate source exhibiting long-term quasi-periodic variability in $\gamma$ rays. A systematic all-sky search for cyclic $\gamma$-ray emission using \emph{Fermi}-LAT data revealed a statistically significant quasi-periodic modulation with a characteristic period of $\sim640$~d, detected at high significance against white noise and remaining significant after accounting for red-noise processes \citep{2017MNRAS.471.3036P}. A refined likelihood analysis yielded a consistent period of $\sim642$~d, significant at $>5\sigma$ against white noise and with only a $\sim0.4$~per~cent probability of arising from a power-law noise spectrum \citep{2017MNRAS.471.3036P}. Over the LAT baseline considered in that study, this corresponds to approximately 4.5 modulation cycles. Importantly, PKS~2052$-$47 belongs to a small group of high-redshift ($z=1.49$) blazars showing indication for quasi-periodic behaviour, making it particularly interesting in the context of long-term jet-modulation or binary supermassive black hole (SMBBH) scenarios.

In this work, we present a comprehensive study of PKS~2052$-$47 based on long-term time-domain analysis of its \emph{Fermi}-LAT $\gamma$-ray light curves. We perform an independent and dedicated search for quasi-periodic oscillation (QPO) signatures, applying rigorous significance tests and explicitly accounting for stochastic variability processes.
We apply multiple time-series techniques—including Lomb--Scargle periodograms (LSP), weighted wavelet $Z$-transforms (WWZ), phase-folding analysis, first-order autoregressive modelling (REDFIT), date-compensated discrete Fourier transform (DCDFT), and damped random walk (DRW) simulations—to assess the consistency and temporal stability of the candidate QPO signals. Our goal is to place the periodicity claims on a statistically sound footing while accounting for red noise, uneven sampling, and finite-duration effects. The results critically test the periodicity hypothesis for PKS~2052$-$47 and provide a general framework for future investigations of similar blazar candidates.

The structure of this paper is as follows: Section~\ref{sec:1} describes the data selection and reduction. Section~\ref{QPO} presents the results of various time-series analyses applied to the $\gamma$-ray light curve. Section~\ref{sig_ev} presents the statistical significance assessment of the detected QPOs. Section~\ref{evo} examines the temporal evolution of the detected QPOs. Finally, Section~\ref{sum} summarizes the findings and discusses their physical implications.

\section{Observations and Data reduction}
\label{sec:1}

\subsection{Fermi-LAT}

The \emph{Fermi} Large Area Telescope (\emph{Fermi}-LAT), aboard the \emph{Fermi} Gamma-ray Space Telescope (formerly GLAST), is a space-based high-energy instrument launched in 2008. With a field of view of $\sim2.3$~sr, the LAT operates predominantly in sky-survey mode, providing complete coverage of the $\sim20$~MeV--500~GeV energy range every $\sim3$~h \citep{2009ApJ...697.1071A}. 
The LAT data were processed using the \texttt{Fermitools} software package (version~2.2.0) distributed by the Fermi Science Support Center (FSSC). Standard event selections and quality cuts were applied following the recommendations in the official LAT analysis documentation.\footnote{\url{https://fermi.gsfc.nasa.gov/ssc/data/analysis/}} Photon events belonging to the SOURCE class (\texttt{evclass=128}, \texttt{evtype=3}) were extracted from a circular region of interest (ROI) of radius $15^{\circ}$ centred on the target, together with a zenith-angle cut of $90^{\circ}$ to suppress contamination from Earth-limb $\gamma$ rays.
Spectral analyses were carried out in the 0.1--300~GeV band, and an independent likelihood analysis was performed with the \texttt{FERMIPY} package (v1.0.1; \citealt{2017ICRC...35..824W}). The Galactic diffuse emission was described using the template \texttt{gll\_iem\_v07.fits}, while the isotropic background component was modeled with \texttt{iso\_P8R3\_SOURCE\_V3\_v1.txt}. The post-launch instrument response function \texttt{P8R3\_SOURCE\_V3} was adopted throughout.
Our source model included all objects from the fourth \emph{Fermi}-LAT catalog (4FGL) located within $25^{\circ}$ of the ROI centre. The normalizations of sources lying within $10^{\circ}$ were allowed to vary during the likelihood fit, whereas more distant sources were fixed at their catalog values. For PKS~2052$-$47, both the flux normalization and spectral-shape parameters ($\alpha$ and $\beta$) were left free. A monthly binned $\gamma$-ray light curve was  generated for the source. The analysis covered the interval MJD~54727.99--58507.99, corresponding to the first $\sim$11~yr of \emph{Fermi}-LAT operations and encompassing the multiple activity states relevant for the QPO search.

\section{Quasi-periodic oscillation}
\label{QPO}
To search for quasi-periodic signatures in the $\gamma$-ray light curve of PKS~2052$-$47, we applied a range of complementary time-series analysis methods. These comprise the LSP, WWZ, REDFIT, DCDFT, and  DRW modelling. In addition, extensive Monte Carlo simulations were carried out to evaluate the statistical significance of the identified features. The methodologies and their associated results are described in detail in the sections that follow.

\begin{figure*}
    \centering
    \includegraphics[width=0.9\textwidth]{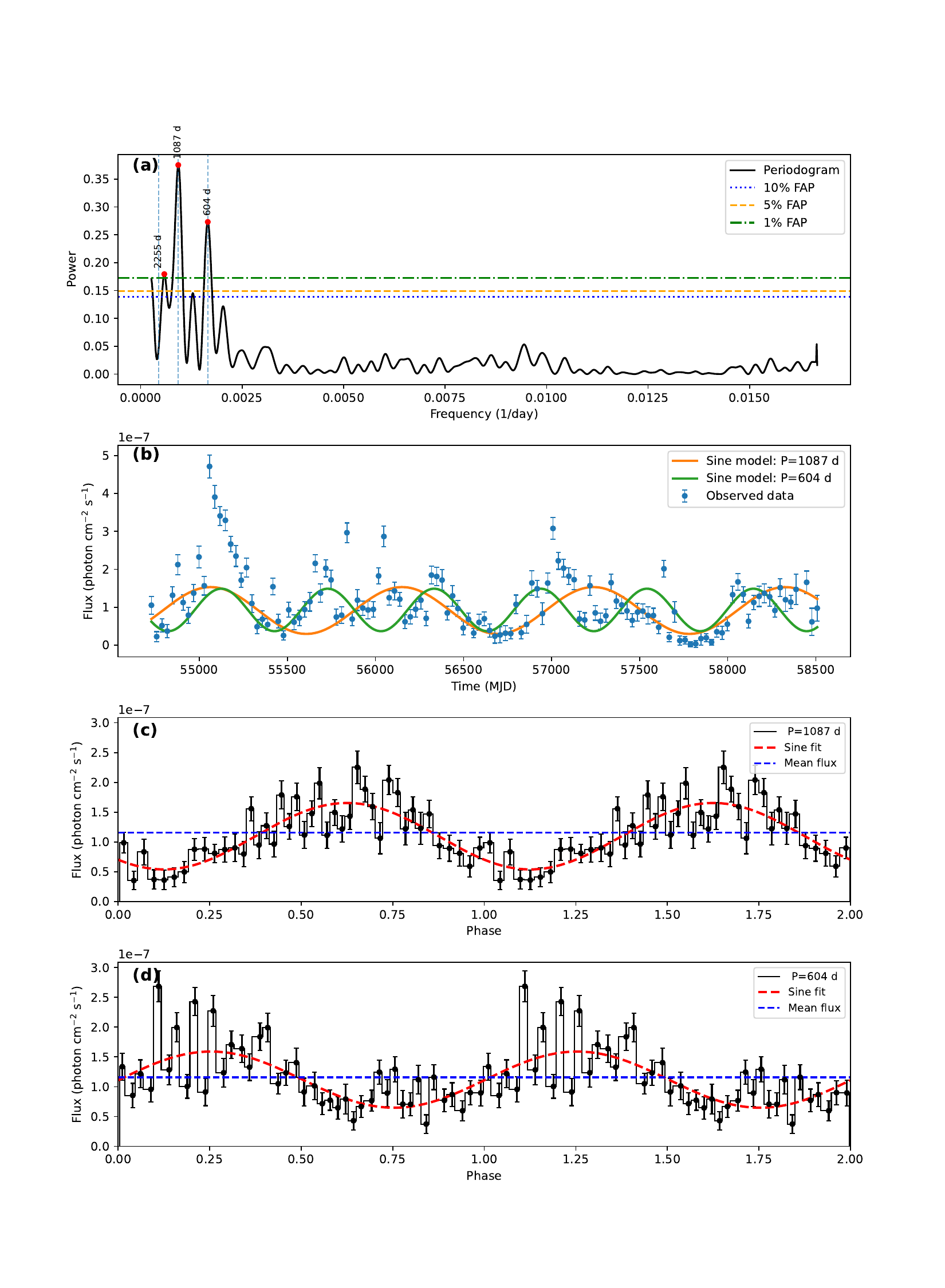}
    \caption{
(a) LSP computed from the monthly binned \emph{Fermi}-LAT data over MJD~54727.99--58507.99. The two most prominent peaks occur at frequencies of $0.000920$ and $0.001655~\mathrm{day^{-1}}$, corresponding to periods of $P_1 = 1087.2 \pm 80.4$~d and $P_2 = 604.2 \pm 23.4$~d, respectively. Horizontal lines mark the false-alarm-probability (FAP) levels of 10\% (blue dotted), 5\% (orange dashed), and 1\% (green dash-dotted). 
(b) $\gamma$-ray light curve overlaid with the best-fitting sinusoidal models corresponding to the two LSP periods. 
(c,d) Phase-folded light curves for $P_1$ and $P_2$, respectively, shown over two full cycles (phase 0–2). In each panel, the data points are plotted in black, the best-fit sinusoidal model is shown by the red dashed curve, and the horizontal blue dashed line indicates the mean flux level. The folding highlights the quasi-periodic modulation present in the $\gamma$-ray emission.
}

    \label{LSP}    
\end{figure*}

\subsection{Lomb-Scargle Periodogram (LSP)}\label{sec:lsp}

The Lomb--Scargle periodogram (LSP) is a widely used technique for identifying periodic signals in unevenly sampled time-series data \citep{lomb1976least,scargle1982studies}. Owing to its suitability for irregularly spaced light curves, it has become a standard tool in astronomical variability analyses. In this work, we employed the \textsc{Astropy} implementation of the Lomb--Scargle algorithm\footnote{\url{https://docs.astropy.org/en/stable/timeseries/lombscargle.html}}, incorporating the measured flux uncertainties to improve the stability of the resulting periodograms. Our application of the LSP follows the approach adopted in our earlier variability studies \citep{zxgv-fzv5,2026arXiv260113181N}. The frequency search was restricted to the range $f_{\rm min}=1/T$ to $f_{\rm max}=1/(2\Delta T)$, where $T$ denotes the total temporal baseline and $\Delta T$ the characteristic sampling interval; further details of the underlying formalism are given by \citet{vanderplas2018understanding}.

The periodogram reveals two prominent peaks at frequencies of $0.000920 \pm 0.000068~\mathrm{d^{-1}}$ and $0.001655 \pm 0.000064~\mathrm{d^{-1}}$, corresponding to periods of $1087.2 \pm 80.4$~d and $604.2 \pm 23.4$~d, respectively (panel~(a) of Figure~\ref{LSP}). The uncertainties on the inferred periods were estimated by fitting Gaussian profiles to the periodogram peaks and adopting the corresponding half-width at half-maximum (HWHM) as the error measure \citep{vanderplas2018understanding,zxgv-fzv5}. To further examine these modulations, we constructed phase-folded $\gamma$-ray light curves using the two LSP periods and fitted them with sinusoidal models. The resulting folded profiles display coherent variations over multiple cycles, supporting the  quasi-periodic nature of the emission. For visual clarity, two full cycles are shown in each panel (see panels~(c) and (d) of Figure~\ref{LSP}).

To further examine the stability of these features, we computed the generalized Lomb--Scargle periodogram (GLSP), which explicitly accounts for measurement uncertainties. The GLSP results are consistent with those obtained from the standard LSP, lending additional support to the presence of these candidate periodicities.
To evaluate the statistical significance of the detected periodicities, we employed the \texttt{LombScargle.false\_alarm\_probability()} routine from the \texttt{astropy.timeseries} module with \texttt{method="baluev"}, which provides an analytic approximation of the false-alarm probability (FAP). This approach is based on the formalism developed by \citet{2008MNRAS.385.1279B}, which uses extreme-value statistics to estimate the FAP while accounting for the effective number of trials across the scanned frequency range. The number of independent frequencies, $N_{\rm ind}$, is internally inferred from the time sampling, frequency grid, and total temporal baseline, and therefore does not need to be specified explicitly. This method yields a statistically grounded estimate of the probability that a given peak in the periodogram arises from stochastic fluctuations, enabling a quantitative assessment of its significance.

\begin{figure*}
    \centering
    \begin{subfigure}[t]{0.95\textwidth}
        \centering
        \includegraphics[width=\textwidth]{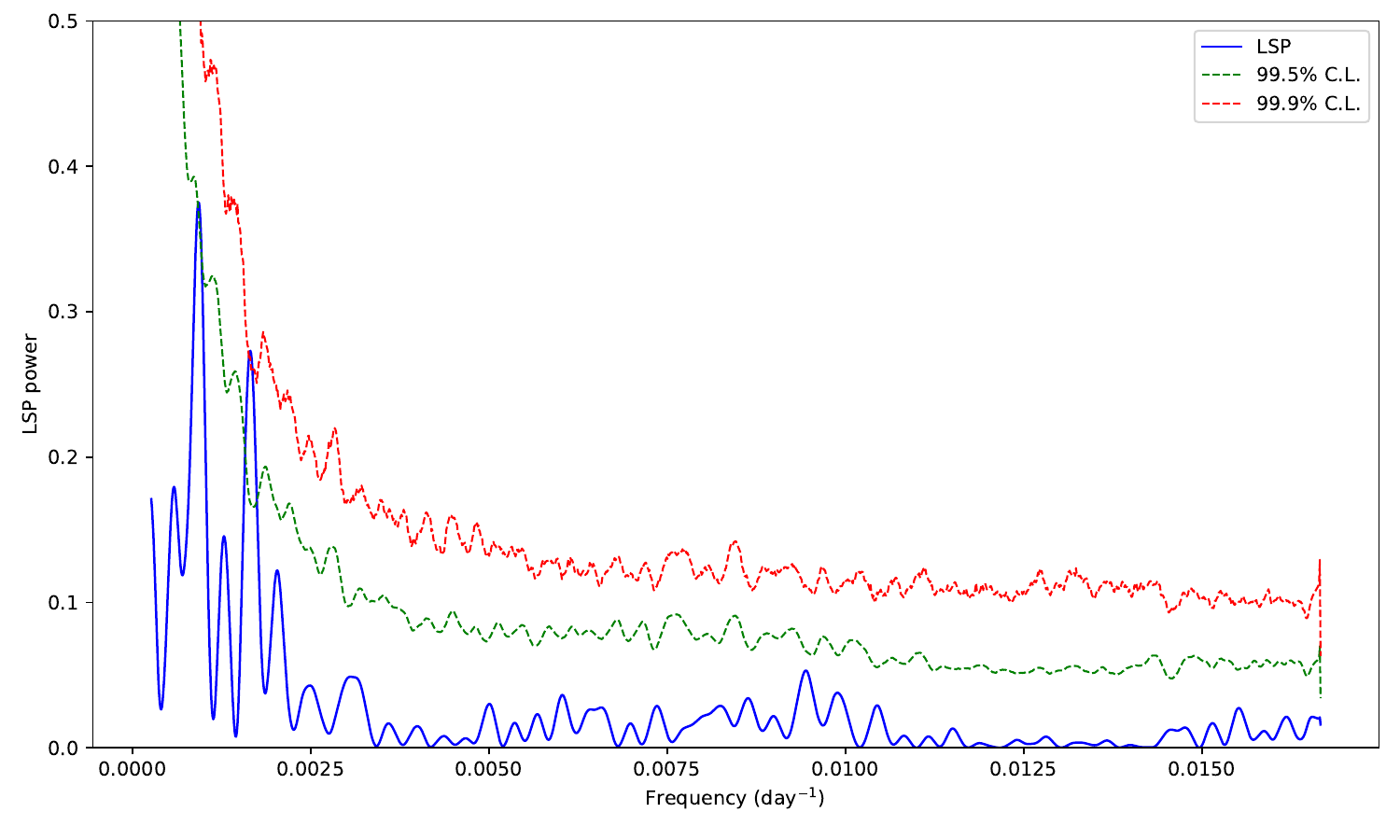}
        \label{sinefit}
    \end{subfigure}
    
    \vspace{0.8em}
    
    \begin{subfigure}[t]{0.95\textwidth}
        \centering
        \includegraphics[width=\textwidth]{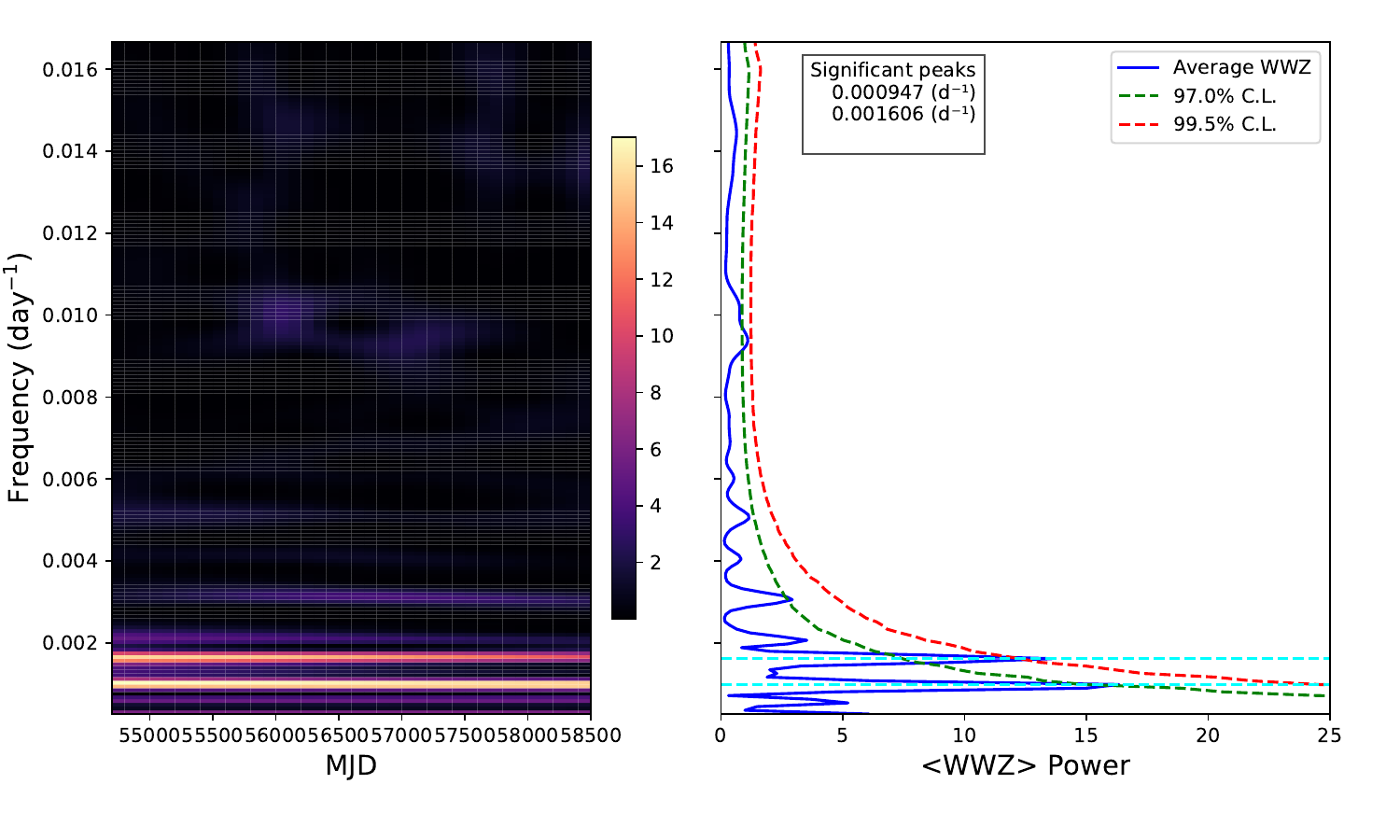}
        \label{LSP_sim}
    \end{subfigure}
    
    \caption{Top: LSP of the monthly binned $\gamma$-ray light curve of PKS~2052$-$47, showing two dominant peaks
at frequencies of $0.000920$ and $0.001655~\mathrm{d^{-1}}$ (periods of $\sim1087$ and $\sim604$~d),
both exceeding the 99.5\% confidence level derived from $10^{5}$ Monte Carlo simulations following
the method of \citet{emmanoulopoulos2013generating}. Bottom: Left: WWZ map of the $\gamma$-ray light curve showing the evolution of power as a function of time (MJD) and frequency. Right: The average WWZ power spectrum with confidence levels derived
from $10^{5}$ Monte Carlo simulations. The dashed green and red lines represent the 97.0\% and
99.5\% confidence levels, respectively. The dotted and dashed cyan lines mark the two most
prominent peaks at frequencies of 0.000947 and 0.001606~day$^{-1}$, corresponding to periods of
$\sim1056$ and $\sim623$~d; the shorter-period feature exceeds the 99.5\% level, while the
longer-timescale peak reaches the $\sim97$\% level.}

    \label{fig:sine_lsp_combined}
\end{figure*}

\subsection{Weighted Wavelet Z-Transform (WWZ)}\label{sec:wwz}

The weighted wavelet $Z$-transform (WWZ; \citealt{foster1996wavelets}) maps an unevenly sampled time series simultaneously into the time and frequency domains by convolving the light curve with a localized oscillatory kernel. This approach is particularly well suited for identifying transient or evolving periodicities, as it provides information on both the characteristic timescale and the epoch over which a given modulation is present. In addition, the WWZ power is expected to diminish when the periodic component becomes weaker, allowing changes in the strength of the signal with time to be tracked.

For the present study, we employed the abbreviated Morlet kernel, defined as
\begin{equation}
f[\omega(t-\tau)] = \exp\!\left[i\omega(t-\tau)-c\,\omega^{2}(t-\tau)^{2}\right],
\end{equation}
with the corresponding WWZ projection
\begin{equation}
W[\omega,\tau:x(t)] = \omega^{1/2}\int x(t)\,f^{*}[\omega(t-\tau)]\,dt,
\end{equation}
where $f^{*}$ is the complex conjugate of the kernel, $\omega$ denotes the angular frequency, and $\tau$ is the temporal offset. The analysis was performed using the publicly available Python implementation of the WWZ algorithm\footnote{\url{https://github.com/eaydin/WWZ}}.

The resulting WWZ power distribution exhibits a dominant peak at $0.001606 \pm 0.000110~\mathrm{d^{-1}}$ and a secondary longer-timescale feature at $0.000947 \pm 0.000100~\mathrm{d^{-1}}$, corresponding to periods of $622.70 \pm 42.94$~d and $1055.62 \pm 112.33$~d, respectively. Uncertainties on the inferred periods were obtained by fitting Gaussian profiles to the peaks in the average WWZ spectrum.


\begin{figure*}
    \centering
    \includegraphics[width=1.0\textwidth]{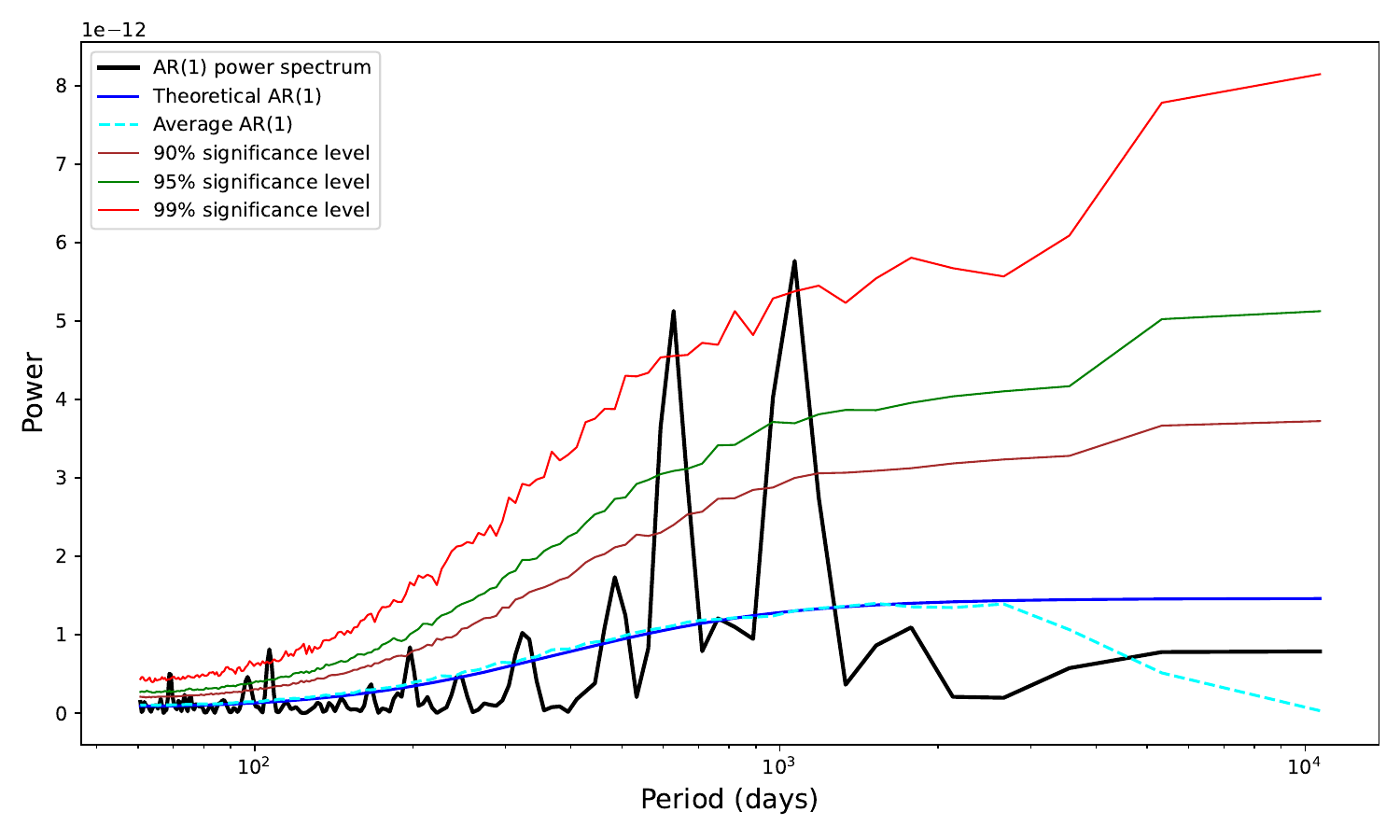}
    \caption{Red-noise-corrected REDFIT power spectrum of the $\gamma$-ray light curve of PKS~2052$-$47. The black curve shows the observed spectrum, while the blue and cyan curves correspond to the theoretical and ensemble-averaged AR(1) models, respectively. The brown, green, and red horizontal lines indicate the 90\%, 95\%, and 99\% Monte Carlo confidence levels. Two dominant peaks are present, both exceeding the 99\% significance threshold.
}
    \label{Fig-redfit}    
\end{figure*}

\subsection{First-order Autoregressive Process (REDFIT)}\label{sec:redfit}

AGN light curves, including those of blazars, are typically characterized by red-noise variability driven by stochastic processes in the jet or accretion disc. Such behaviour can be represented by a first-order autoregressive [AR(1)] model, in which the emission at a given epoch depends linearly on the immediately preceding value \citep{schulz2002redfit}. This can be expressed as $r(t_i)=A\,r(t_{i-1})+\epsilon(t_i)$, where $A=\exp[-(t_i-t_{i-1})/\tau]$ is the autoregressive coefficient associated with the characteristic timescale $\tau$, and $\epsilon(t_i)$ denotes a random driving term. The corresponding theoretical power spectrum takes the form
\begin{equation}\label{redfiteq}
G_{rr}(f_i)=G_0\,\frac{1-A^{2}}{1-2A\cos\!\left(\pi f_i/f_{\rm Nyq}\right)+A^{2}},
\end{equation}
where $G_0$ is the mean spectral power, $f_i$ are the temporal frequencies, and $f_{\rm Nyq}$ denotes the Nyquist frequency.

The red-noise-corrected spectrum was computed using the REDFIT routine implemented in \texttt{R}\footnote{\url{https://rdrr.io/cran/dplR/man/redfit.html}}, and the statistical significance of candidate features was assessed via Equation~(\ref{redfiteq}). This procedure reveals two prominent peaks at $0.000937 \pm 0.000101~\mathrm{d^{-1}}$ and $0.00159 \pm 0.00012~\mathrm{d^{-1}}$, corresponding to periods of approximately 1067 and 628~d, respectively; both exceed the 99\% confidence level (see Figure~\ref{Fig-redfit}). The uncertainties in the peak frequencies were estimated from the half-width at half-maximum (HWHM) of Gaussian fits to the REDFIT spectrum.

\begin{figure*}
    \centering
    \includegraphics[width=0.95\textwidth]{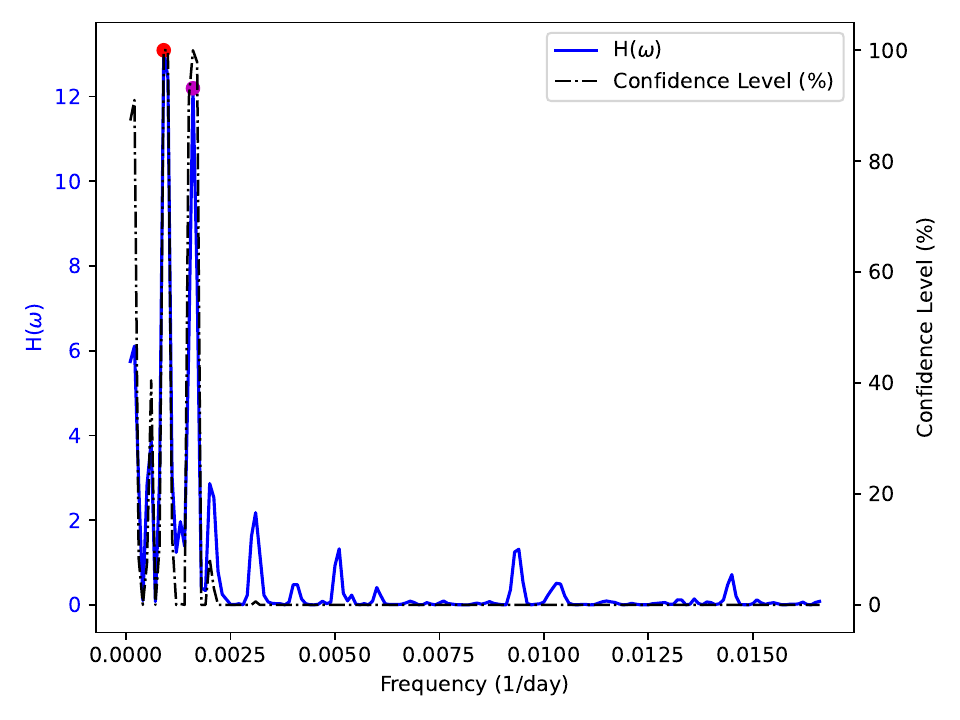}
    \caption{ Modified periodogram $H(\omega)$ obtained using the DCDFT method for the $\gamma$-ray light curve of PKS~2052$-$47. Two dominant peaks are detected at periods of $\sim1111$ and $\sim625$~d, consistent with the timescales inferred from the other timing techniques. The associated uncertainties were estimated by fitting Gaussian profiles to the corresponding features in the periodogram. The black dash--dotted curve denotes the 99\% confidence level, which is exceeded by both peaks, indicating that they are statistically significant.}

    \label{Fig-DCDFT}    
\end{figure*}

\subsection{Date-compensated Discrete Fourier Transform (DCDFT)}\label{sec:dcdft}
Estimating power spectra from unevenly sampled time series remains a central difficulty in the search for quasi-periodic oscillations in blazar light curves \citep[e.g.,][]{fan2007radio}. When applied to such data, the classical discrete Fourier transform (DFT) can suffer from frequency shifts and amplitude distortions, which may compromise the reliability of detected periodic components. These effects are alleviated by the date-compensated discrete Fourier transform (DCDFT; \citealt{ferraz1981estimation,foster1995cleanest}), which models the light curve through a least-squares fit that includes both sinusoidal and constant terms, rather than relying exclusively on purely sinusoidal components as in the standard DFT. This formulation is particularly advantageous at low frequencies ($<0.02~\mathrm{day^{-1}}$), where uneven sampling can otherwise introduce discrepancies of several per cent.

In our analysis, we applied the DCDFT to the $\gamma$-ray light curve using publicly available software\footnote{\url{https://github.com/ilmimris/dcdft}}, which follows the formalism described by \citet{ferraz1981estimation}. The resulting modified periodogram $H(\omega)$ exhibits two prominent peaks at $0.00090 \pm 0.00015~\mathrm{d^{-1}}$ and $0.00160 \pm 0.00020~\mathrm{d^{-1}}$, corresponding to periods of $\sim1111$ and $\sim625$~d, respectively. Both features exceed the 99\% confidence level (see Figure~\ref{Fig-DCDFT}), indicating that they are statistically significant. Because the frequency grid in the DCDFT remains uniform regardless of data gaps, the method is well suited to long-term, irregularly sampled blazar monitoring.

The periodicities inferred from all four techniques are mutually consistent within uncertainties, with the shorter timescale emerging as the dominant modulation, as summarized in Table~\ref{tab:QPO_all}.

In this study, we report indication for a dominant quasi-periodic modulation in the $\gamma$-ray emission of PKS~2052$-$47 at a characteristic period of $\sim604$~d, together with a secondary longer-timescale feature near $\sim1087$~d inferred from the LSP. The shorter of the two timescales is consistently recovered by the WWZ, REDFIT, and DCDFT analyses, which yield periods in the range $\sim620$--628~d, in close agreement with the $\sim642$~d modulation previously reported from weekly binned data by \citep{2017MNRAS.471.3036P}.

\begin{table*}
\setlength{\extrarowheight}{6pt}
\setlength{\tabcolsep}{6pt}
\centering
\caption{Summary of QPO frequencies detected for PKS~2052$-$47 using different timing techniques.
Uncertainties are quoted where available, and the local significance levels are given in parentheses.
Frequencies are expressed in units of $10^{-3}~\mathrm{day^{-1}}$.}

\resizebox{0.95\textwidth}{!}{%
\begin{tabular}{c c c c c}
\hline
4FGL Name  
& LSP $\left(10^{-3}~\mathrm{day^{-1}}\right)$ 
& WWZ $\left(10^{-3}~\mathrm{day^{-1}}\right)$ 
& REDFIT $\left(10^{-3}~\mathrm{day^{-1}}\right)$ 
& DCDFT $\left(10^{-3}~\mathrm{day^{-1}}\right)$ \\
\hline
4FGL~J2056.2$-$4714 
& 0.920$\pm$0.068 ($>99.5\%$) 
& 0.947$\pm$0.100 ($>97\%$) 
& 0.937$\pm$0.101 ($>99\%$) 
& 0.900$\pm$0.150 ($>99\%$) \\
& 1.655$\pm$0.064 ($>99.9\%$) 
& 1.606$\pm$0.110 ($>99.5\%$) 
& 1.590$\pm$0.120 ($>99\%$) 
& 1.600$\pm$0.200 ($>99\%$) \\
\hline
\end{tabular}
}
\label{tab:QPO_all}
\end{table*}

\subsection{Damped Random Walk Model}
\label{sec:drw}

The observed variability in AGN emission is stochastic in nature and can be well described by the simplest form of a continuous autoregressive moving average process, CARMA(1,0) \citep[e.g.,][]{moreno2019stochastic, burke2021characteristic, zhang2022characterizing, zhang2023gaussian, sharma2024microquasars,TANTRY2025100372}. This formulation is commonly referred to as the damped random walk (DRW) model and has been widely adopted to characterize red-noise variability in AGN light curves.

In this framework, the flux variations are governed by the stochastic differential equation

\begin{equation}
\left[\frac{d}{dt} + \frac{1}{\tau_{\rm DRW}}\right] y(t)
= \sigma_{\rm DRW}\,\epsilon(t),
\end{equation}

where $\tau_{\rm DRW}$ denotes the characteristic damping timescale, $\sigma_{\rm DRW}$ is the long-term variability amplitude, and $\epsilon(t)$ is a Gaussian white-noise process with zero mean and unit variance. The covariance function of the DRW process is given by

\begin{equation}
k(\Delta t) = \sigma_{\rm DRW}^{2}\,
\exp\!\left(-\frac{|\Delta t|}{\tau_{\rm DRW}}\right),
\end{equation}

where $\Delta t = |t_n - t_m|$ represents the time lag between two measurements.

The power spectral density (PSD) associated with the DRW model has the form of a broken power law,

\begin{equation}
S(\omega) =
\sqrt{\frac{2}{\pi}}\,
\frac{\sigma_{\rm DRW}^{2}\,\tau_{\rm DRW}}
{1 + (\omega\,\tau_{\rm DRW})^{2}},
\end{equation}

where $\omega$ is the angular frequency. At low frequencies ($\omega \ll 1/\tau_{\rm DRW}$), the PSD is approximately flat, while at high frequencies ($\omega \gg 1/\tau_{\rm DRW}$) it asymptotically approaches a $P(\omega)\propto \omega^{-2}$ form, characteristic of red-noise dominated processes.

\begin{figure*}
    \centering
    \includegraphics[width=0.48\textwidth]{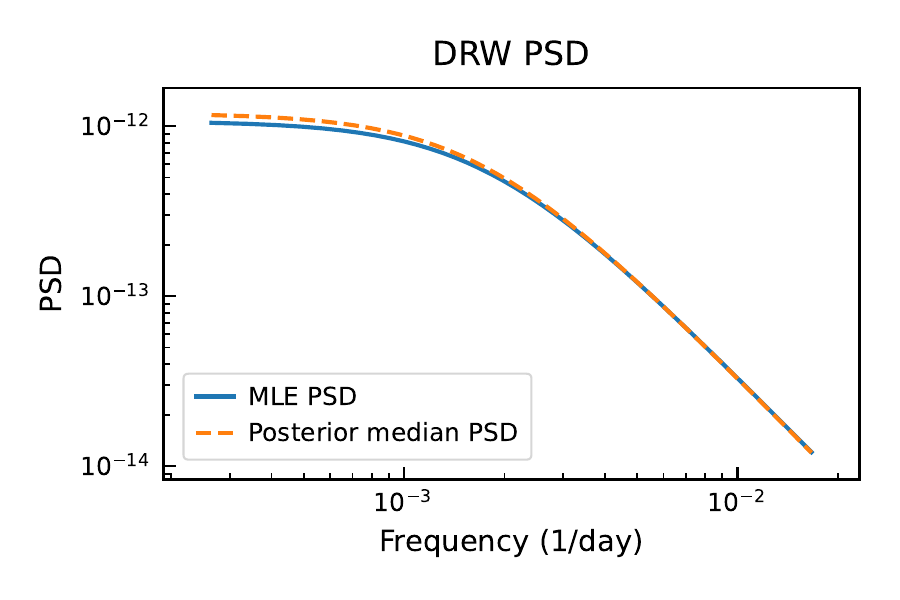}
    \hfill
    \includegraphics[width=0.48\textwidth]{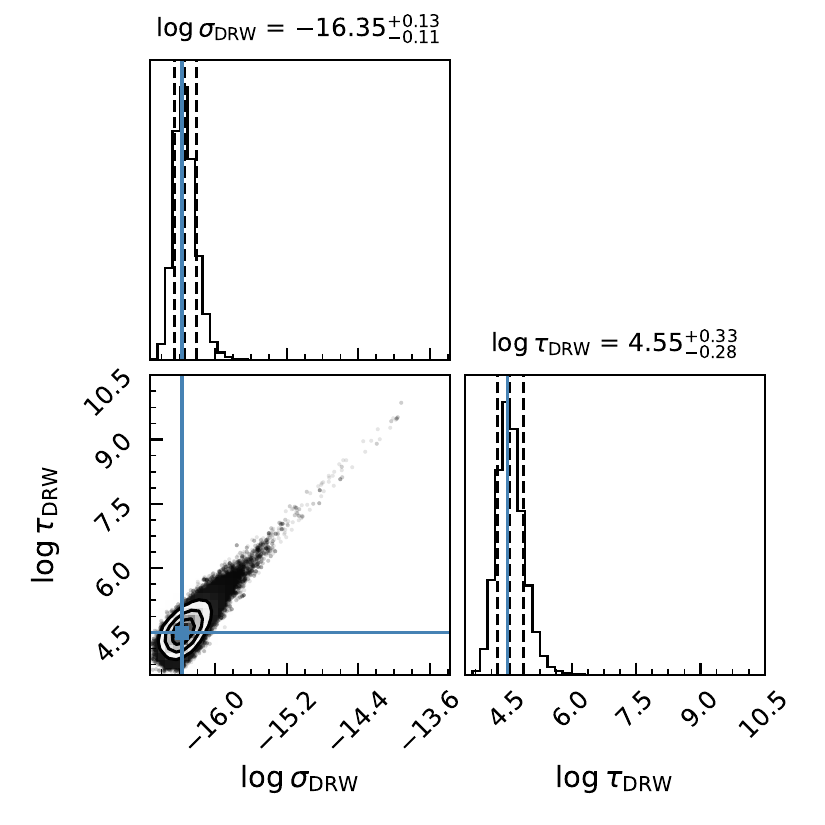}
    \caption{
    PSDs derived from the DRW modeling of the $\gamma$-ray light curve.
    {\it Left:} PSD obtained using the maximum-likelihood DRW parameters together with the posterior-median solution.
    {\it Right:} Corresponding PSD representation derived from the MCMC posterior samples.
    Both panels exhibit the characteristic low-frequency flattening and a steep high-frequency decline expected for a DRW process.
    The inferred parameters are $\ln\sigma_{\rm DRW}=-16.35$ and
$\ln\tau_{\rm DRW}=4.55^{+0.33}_{-0.28}$ (days), consistent with the maximum-likelihood values
$\ln\sigma_{\rm DRW}=-16.37$ and $\ln\tau_{\rm DRW}=4.49$.}

    \label{fig:drw_psd_mle}
\end{figure*}

We modeled the $\gamma$-ray light curve of PKS~2052$-$47 using the publicly available \textsc{EzTao} package\footnote{\url{https://eztao.readthedocs.io/en/latest/}}, which is built on top of the \textsc{celerite} Gaussian-process framework\footnote{\url{https://celerite.readthedocs.io/en/stable/}} \citep{2013PASP..125..306F}. Parameter estimation was performed via Markov Chain Monte Carlo (MCMC) sampling implemented using the \texttt{emcee} package\footnote{\url{https://github.com/dfm/emcee}} within \textsc{EzTao}. In total, 25\,000 MCMC realizations were generated, with the first 10\,000 steps discarded as burn-in to ensure convergence of the chains. From the remaining samples, we derived the posterior distributions of the DRW model parameters and their associated uncertainties.
The DRW power spectral densities derived from the maximum-likelihood solution and from the posterior-median parameters are shown in Figure~\ref{fig:drw_psd_mle}. The solid curve corresponds to the PSD computed from the maximum-likelihood DRW parameters, while the dashed curve shows the PSD obtained using the posterior-median values from the MCMC sampling. Both spectra exhibit the characteristic low-frequency flattening and a steep high-frequency decline, consistent with the broken-power-law behaviour expected for a DRW process, with a break near $f\sim1/\tau_{\rm DRW}$. This feature corresponds to a characteristic stochastic timescale of $\tau_{\rm DRW}\approx90$~d. The quasi-periodic modulation detected at $\sim600$--630~d lies well beyond this timescale, indicating that it cannot be attributed to intrinsic DRW variability and instead requires an additional, non-stochastic component.
The MCMC analysis yields median values of $\ln\sigma_{\rm DRW}=-16.35$ and
$\ln\tau_{\rm DRW}=4.55^{+0.33}_{-0.28}$ (with $\tau_{\rm DRW}$ in days), in excellent agreement
with the maximum-likelihood estimates of $\ln\sigma_{\rm DRW}=-16.37$ and
$\ln\tau_{\rm DRW}=4.49$. This close correspondence demonstrates that the DRW model provides a
stable and statistically well-constrained description of the stochastic variability in the
$\gamma$-ray light curve.

\section{Significance Evaluation}
\label{sig_ev}

\begin{figure*}
    \centering
    \includegraphics[width=1.0\textwidth]{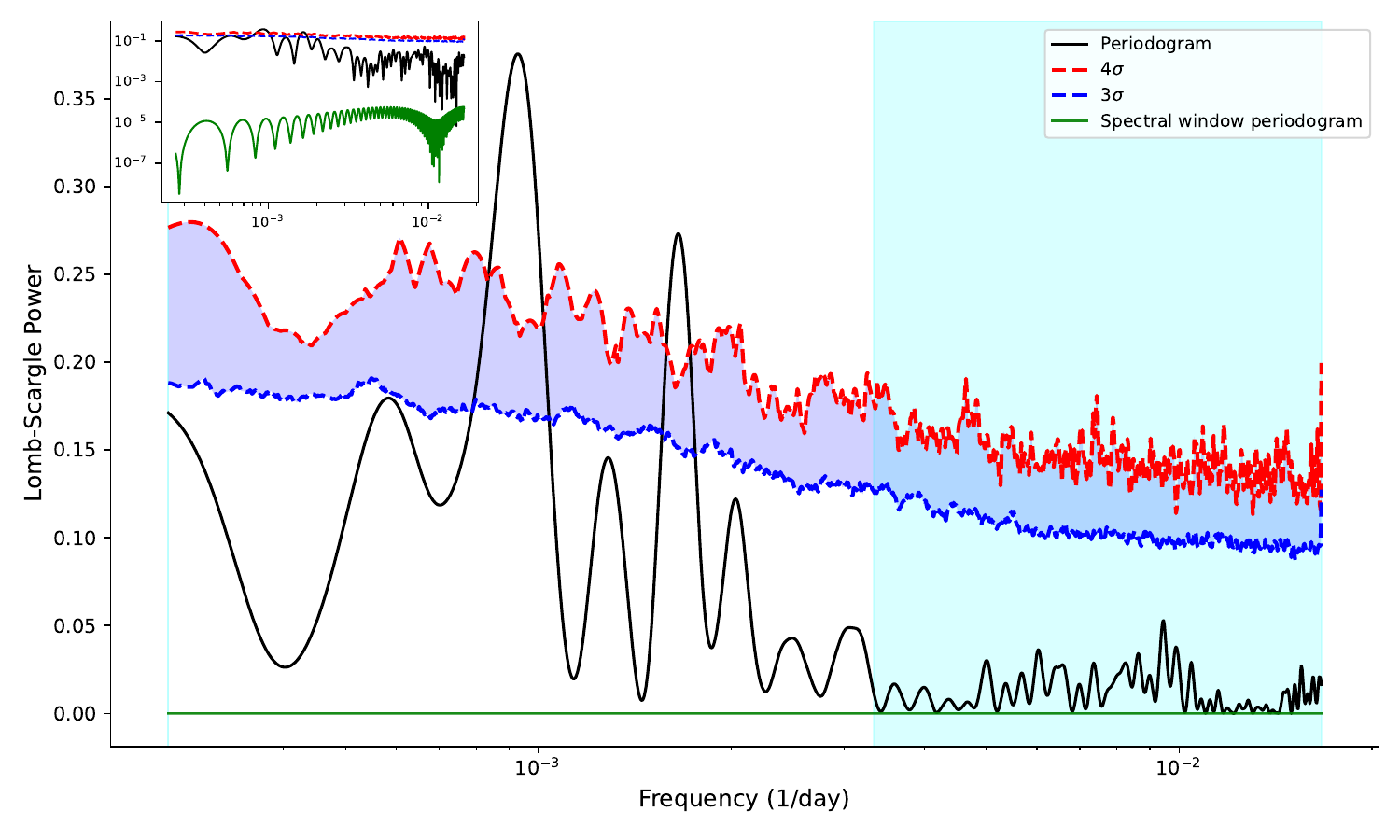}
    \caption{LSP  of the monthly binned $\gamma$-ray light curve of PKS~2052$-$47 (black curve), together with the spectral-window periodogram (green) constructed to assess the effects of uneven temporal sampling. The dashed red curve shows the $4\sigma$ significance threshold derived from 20\,000 DRW simulations. Two dominant peaks near $\sim0.000920$ and $\sim0.001655~\mathrm{day^{-1}}$ exceed this threshold, corresponding to periods of $\sim1087$ and $\sim604$~d, respectively. The cyan-shaded region marks the frequency range deemed unreliable owing to the finite duration of the light curve and cadence-based criteria adopted following \citep{TANTRY2025100372}. The inset panel presents the same periodograms on logarithmic axes.
}
    \label{Fig-dr}    
\end{figure*}

The red-noise variability commonly observed in AGN and blazar light curves is generally attributed to stochastic processes in the jet or accretion flow and can be described by a power-law power spectral density (PSD) of the form $P(\nu)\propto A\nu^{-\beta}$, where $\nu$ denotes the temporal frequency and $\beta>0$ is the spectral index. For PKS~2052$-$47, \citet{2017MNRAS.471.3036P} reported a best-fitting PSD slope of $\beta = 0.69$.
We adopted a Monte Carlo approach to quantify the statistical significance of the dominant peaks detected in the LSP and WWZ analyses. Specifically, $10^{5}$ synthetic light curves were generated that reproduce both the power spectral density (PSD) and probability density function (PDF) of the observed data, following the prescription of \citet{emmanoulopoulos2013generating}. The local significance of each candidate feature was estimated from the distribution of spectral powers at the corresponding frequencies across the simulated realizations.
Within the LSP analysis, the two principal peaks at $\sim0.000920~\mathrm{d^{-1}}$ ($\sim1087$~d) and $\sim0.001655~\mathrm{d^{-1}}$ ($\sim604$~d) both exceed the 99.5\% confidence level. Consistent signals are recovered in the WWZ analysis, with peak frequencies clustered around $\sim0.001606~\mathrm{d^{-1}}$ and $\sim0.000947~\mathrm{d^{-1}}$. The shorter-period signal exceeds the 99.5\% confidence level, while the longer-timescale feature reaches the $\sim97$\% confidence
level in the average WWZ spectrum. The REDFIT results further support these detections, yielding dominant features near $\sim 0.000937 ~\mathrm{d^{-1}}$ and $\sim0.0016~\mathrm{d^{-1}}$ with significance levels greater than 99\%.
Taken together, these findings provide indication for transient quasi-periodic modulations in the $\gamma$-ray light curve of PKS~2052$-$47 over the interval MJD~54727.99--58507.99, corresponding to approximately four and six modulation cycles for the longer and shorter timescales, respectively (see the  panel (b) of Figure~\ref{fig:sine_lsp_combined}).
In addition to the significance tests described above, we examined the $\gamma$-ray light curve using DRW framework, which is widely employed to characterize red-noise variability in AGN emission. The DRW-based significance testing and spectral-window analysis adopted here follow the methodology described in detail by \citep{TANTRY2025100372}; we therefore provide only a concise summary and emphasize the elements specific to the present study.
Using the \textsc{EzTao} package, we generated 20,000 mock $\gamma$-ray light curves based on the maximum-likelihood DRW parameters inferred from the data and sampled them in an identical manner to the observed light curve. LSP were computed for each realization, and percentile thresholds corresponding to the $1\sigma$, $2\sigma$, $3\sigma$, and $4\sigma$ confidence levels were derived at each trial frequency from the ensemble of simulations. We further constructed spectral-window periodograms to evaluate the impact of uneven sampling, following the procedure outlined by \citep{TANTRY2025100372}. In this DRW-based framework, two prominent features near $\sim0.000920~\mathrm{day^{-1}}$ and $\sim0.001655~\mathrm{day^{-1}}$—corresponding to periods of $\sim1087$ and $\sim604$~d—exceed the $4\sigma$ confidence envelope. The corresponding DRW significance curves and spectral-window periodograms are shown in Figure~\ref{Fig-dr}.

\begin{figure*}
    \centering
    \includegraphics[width=0.95\textwidth]{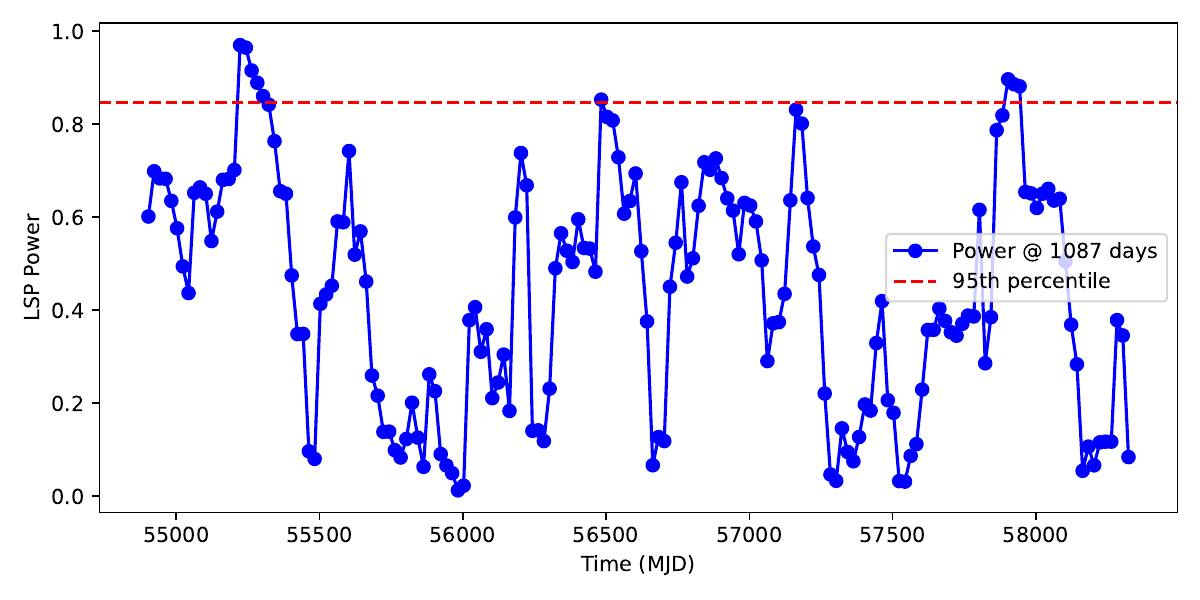}
    \includegraphics[width=0.95\textwidth]{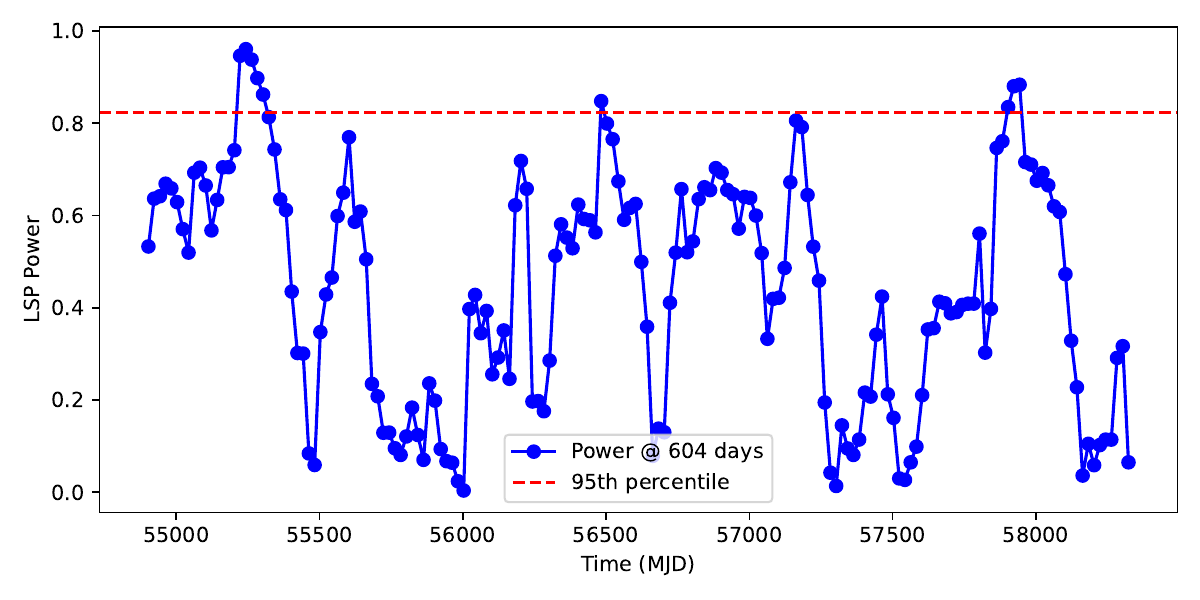}
    \caption{Windowed Lomb--Scargle power as a function of time at fixed periods of 1087.2 and 604.2~d. A sliding-window approach is used to trace the temporal evolution of the quasi-periodic signals across the full observational baseline. The horizontal dashed line marks the 95th-percentile significance threshold. Epochs in which the power exceeds this level indicate that the QPOs are intermittent and do not persist uniformly throughout the light curve.}

    \label{Fig-window}    
\end{figure*}

\section{Temporal Evolution of the QPO}
\label{evo}
To assess the temporal persistence of the candidate QPOs at periods of 1087.2 and 604.2~d, we performed a sliding-window Lomb--Scargle analysis. In this method, the light curve was partitioned into overlapping temporal segments, within which the LSP power was evaluated at the two fixed periods. The resulting evolution of the LSP power, presented in Figure~\ref{Fig-window}, shows that the strength of the signal changes over time. During several intervals the power rises above the 95th-percentile level, indicating that the oscillations are not present throughout the full data set but instead emerge episodically. This behaviour is typical of non-stationary or evolving QPOs and is compatible with scenarios in which the underlying physical process operates intermittently, potentially linked to time-variable conditions in the jet or accretion environment.

\section{Summary and Discussion}
\label{sum}

 We conducted a systematic search for long-timescale QPOs in the $\gamma$-ray emission of the high-redshift FSRQ PKS~2052$-$47 using monthly binned \emph{Fermi}-LAT data covering MJD~54727.99--58507.99. To minimise method-dependent biases and to account for uneven sampling and red-noise variability, we applied multiple, complementary timing techniques, including  LSP, WWZ, REDFIT, and  DCDFT.
 Our principal findings can be summarised as follows.
\begin{itemize}
\item Taken together, the multi-technique timing analyses reveal one dominant quasi-periodic modulation on a timescale of $\sim600$--630~d and a secondary longer-timescale feature near $\sim1050$--1110~d in the $\gamma$-ray light curve of PKS~2052$-$47.

\item The LSP reveals prominent peaks at $f\simeq9.20\times10^{-4}\,\mathrm{d^{-1}}$ and $f\simeq1.66\times10^{-3}\,\mathrm{d^{-1}}$, corresponding to $P_{1}\simeq1087$~d and $P_{2}\simeq604$~d. Phase-folded light curves at both periods show coherent modulations consistent with quasi-sinusoidal variability.
\item The WWZ analysis independently detects signals at frequencies consistent with the LSP peaks and localises their power in time, indicating an evolving, non-stationary periodic component rather than a strictly persistent oscillation.
\item Emmanoulopoulos-type Monte Carlo simulations that reproduce both the PSD and PDF of the observed
light curve demonstrate that both candidate periods exceed the $\gtrsim99.5$~per~cent local
confidence levels in the LSP spectrum, while in the WWZ analysis the shorter-period signal exceeds
this threshold and the longer-timescale feature reaches the $\sim97$~per~cent confidence level.

\item Independent red-noise modelling with a DRW framework yields a characteristic stochastic timescale of $\tau_{\rm DRW}\sim90$~d. DRW-based simulations based on 20\,000 realisations confirm that the QPO peaks exceed the $4\sigma$ confidence envelope.
\item Spectral-window diagnostics show that neither of the detected periodicities is produced by the sampling pattern of the data.
\item A sliding-window LSP evaluated at fixed $P_{1}$ and $P_{2}$ demonstrates that the QPO power varies with time, with several intervals exceeding the 95th-percentile threshold and others falling below it, implying that the modulation is episodic rather than persistent across the full $\sim11$-yr baseline.
\item Over the $\sim11$~yr baseline, the number of observed cycles is approximately four for the longer timescale and six for the shorter.

\end{itemize}

\subsection{Physical interpretation}

Long-timescale QPOs in blazars continue to attract considerable
interest because they may encode information about the dynamics of the central engine and the
geometry of the relativistic jet. A wide range of physical scenarios has been proposed to explain
year-scale or multi-year modulations in blazar light curves, including 
SMBBH systems, precession or helical motion of jets driven by geometric effects, and
instabilities in the accretion flow that modulate plasma injection into the jet
\citep[e.g.][]{sillanpaa1988oj,xie2008periodicity,valtonen2008massive,rieger2004geometrical,
gupta2008periodic,sandrinelli2016quasi,otero2020quasi,gong2022quasiperiodic,gong2024detection}.
Within this broader context, PKS~2052$-$47 represents a valuable new case study, as a dominant timescale near $\sim600$--630~d and a secondary longer-timescale modulation near $\sim1050$--1110~d are recovered independently by several timing techniques and are found to be statistically significant when tested against realistic red-noise models.

The coexistence of two characteristic periods, together with their episodic appearance in the
sliding-window analysis, provides important clues to the underlying mechanism. In particular, the
intermittency argues against a strictly coherent clock operating throughout the entire observing
baseline and instead favours scenarios in which the physical driver becomes active only during
specific epochs. Such behaviour is naturally produced in geometric models in which the jet
orientation or the trajectory of a compact emitting region evolves quasi-periodically, thereby
modulating the Doppler factor seen by the observer
\citep[e.g.][]{rieger2004geometrical,villata1999helical,otero2020quasi,sharma2024detection}.
Even modest variations in the viewing angle can lead to pronounced flux swings in highly beamed
sources, making Doppler-geometry effects particularly attractive for explaining long-term QPOs in
$\gamma$ rays.

Several specific geometric mechanisms may operate in PKS~2052$-$47. Jet precession driven by
Lense--Thirring torques acting on a misaligned inner accretion flow provides one physically
motivated channel for producing quasi-periodic changes in orientation
\citep[e.g.][]{rieger2004geometrical,gupta2008periodic}. On longer timescales, torques exerted by a
secondary black hole in a close SMBBH system can induce precession or nutation of the disc–jet
system, leading to non-ballistic helical jet trajectories and periodic Doppler modulation
\citep[e.g.][]{sillanpaa1988oj,valtonen2008massive,gong2022quasiperiodic,gong2024detection}.
Alternatively, the emitting plasma may follow a rotating or helical path within a bent or twisted
jet, giving rise to the classical ``lighthouse'' effect discussed in many earlier works
\citep[e.g.][]{villata1999helical,rieger2004geometrical,otero2020quasi,gong2023two}.

In these geometric pictures, the detection of two periods can be interpreted in several ways.
One possibility is that the shorter timescale reflects a fundamental precession or pattern
period, while the longer modulation corresponds to a commensurate or higher-order harmonic
activated during particular epochs. Another possibility is that the two timescales arise from
distinct but coupled processes, such as simultaneous precession and nutation of the jet, thereby
producing a beat-like envelope in the observed flux modulation. Similar near-harmonic structures
have been reported in other blazars, where multiple year-scale periods were interpreted in terms
of orbiting hot spots in the inner accretion disc or oscillatory modes triggered by
Kelvin--Helmholtz instabilities that propagate into the jet
\citep[e.g.][]{an2013periodic,otero2020quasi,gong2022quasiperiodic}. Although the ratio between
$P_{2}\sim600$--630~d and the longer modulation near $\sim1050$--1110~d is not an exact integer, the
finite time baseline, red-noise contamination, and associated uncertainties leave open the
possibility of a near-resonant relationship. The fact that the $\sim600$--630~d modulation is
recovered by all methods, whereas the longer-timescale feature appears with lower or
method-dependent significance, further supports this cautious interpretation.

Accretion-driven interpretations provide a complementary class of models and cannot be ruled out
on the basis of $\gamma$-ray data alone. In such scenarios, oscillations in the innermost disc,
magnetic-flux accumulation or reconnection cycles, or disc–wind interactions may modulate the rate
at which energetic particles are injected into the jet, thereby imprinting quasi-periodic
signatures that are subsequently amplified by relativistic beaming
\citep[e.g.][]{gupta2008periodic,bhatta2016detection,tavani2018blazar,sandrinelli2016quasi}.
Because multiple instability modes or emission zones may operate simultaneously, such intrinsic
processes could also give rise to more than one characteristic timescale, consistent with the two
periodicities reported here.

\textbf{Despite the high statistical significance of the shorter $\sim$600--630\,d period under red-noise-aware tests, and the
more tentative nature of the longer modulation, their interpretation must be tempered by the finite observational baseline and the
limited number of cycles sampled, particularly for the longer timescale. Similar cautions have been raised in several studies of
quasi-periodic variability in blazars. For instance, \citep{sandrinelli2016quasi} reported a $\sim$317\,d quasi-periodic modulation in the optical/NIR emission of the blazar PKS~2155$-$304 together with a possible $\gamma$-ray counterpart near $\sim$642\,d. They noted,
however, that the number of observed cycles was limited and that the stability of the signal over long timescales remains uncertain. Moreover, the authors emphasized that periodicity searches in AGN light curves are complicated by red-noise--dominated variability, irregular sampling, and strong flaring activity, which can introduce apparent periodic features. Continued monitoring is therefore required to establish whether such signals correspond to persistent quasi-periodic modulation. Likewise, \citep{otero2020quasi}
reported year-scale quasi-periodic variability in the optical light curves of the blazars 3C~66A and B2~1633$+$38, with characteristic periods of $\sim$3\,yr and $\sim$1.9\,yr, respectively. Their analysis further indicated that the apparent periodicity in 3C~66A may vary between different activity states of the source, suggesting that such signals can be intermittent rather than strictly persistent. More generally, \citep{gong2022quasiperiodic} carried out a systematic search for quasi-periodic modulation in long-term \textit{Fermi}-LAT blazar light curves and emphasized that red-noise-dominated variability can produce apparent periodic features when only a limited number of cycles are observed. They therefore stressed that careful red-noise modeling and continued long-term monitoring are essential for establishing the persistence of candidate QPO signals. In this context, the episodic appearance of the candidate modulations in PKS~2052$-$47, together with the modest number of observed cycles---especially for the longer $\sim$1050--1110\,d feature---motivates extending the temporal baseline and testing for recurrence with future \textit{Fermi}-LAT observations.}

Looking ahead, a natural next step will be contemporaneous broadband SED modelling, which can constrain key physical parameters of the emission region, including
the Doppler factor, characteristic electron energies, and source geometry. Such information is
essential for evaluating specific geometric or binary-driven scenarios and for translating the
observed QPO timescales into physically meaningful quantities. Complementary radio very-long-
baseline interferometry could test for periodic changes in jet position angle or recurrent
component ejections, while optical polarimetric monitoring may uncover systematic electric-vector
position-angle rotations associated with orientation changes in the emitting region. Together,
these multiwavelength diagnostics would strongly discriminate between Doppler-geometry models and
intrinsic dissipation scenarios and would ultimately enable attempts to connect the observed
timescales to SMBBH-driven dynamics and to derive physically motivated constraints on the
black-hole mass.

Although year-like QPOs in several blazars have been interpreted in terms of SMBBH-driven dynamics
and used to place constraints on black-hole masses in the literature, we do not attempt such
estimates here. Reliable mass inference requires adopting a specific physical model and
independent constraints on the Doppler factor, viewing geometry, and binary mass ratio, which are
beyond the scope of the present timing-focused analysis.

\subsection{Conclusions}
\label{sec:conclusions}

Taken together, the results presented here identify PKS~2052$-$47 as a promising new member of the
small but growing class of blazars exhibiting long-timescale $\gamma$-ray QPOs. We find indication for
a dominant quasi-periodic modulation on a timescale of $\sim600$--630~d, together with a secondary
longer-timescale feature near $\sim1050$--1110~d. Their intermittent behaviour in sliding-window
analyses, and their possible harmonic or near-resonant relationship, provide strong motivation for
continued monitoring and for deeper theoretical modelling aimed at uncovering the physical origin
of these oscillations.

Future progress will require coordinated multiwavelength campaigns, in particular contemporaneous
broadband SED modelling to constrain key physical parameters of the emission region, together with
radio VLBI and optical polarimetric observations to test Doppler-geometry scenarios and possible
SMBBH-driven dynamics. Such efforts will be essential for establishing whether the detected
modulations recur, for quantifying their duty cycle, and for determining whether they arise from
geometric effects in the relativistic jet or from accretion-related instabilities at the jet base.

\section{Acknowledgements}
SAD express  gratitude to the Inter-University Centre for Astronomy and Astrophysics (IUCAA) in Pune, India, for the support and facilities provided.

\bibliographystyle{elsarticle-harv} 
\bibliography{sample631}





 
 \section{Appendix}

\end{document}